\documentclass[a4paper,12pt]{article}
\usepackage{geometry}
\usepackage{graphicx}
\usepackage{array}
\usepackage{hyperref}
\usepackage{amsmath, amssymb}
\usepackage{authblk}
\usepackage{setspace}
\usepackage[labelfont=bf]{caption}
\usepackage{cleveref} 

\title{\textbf{GRiNS: A Python Library for Simulating Gene Regulatory Network Dynamics}}

\author[1]{Pradyumna Harlapur}
\author[2]{Harshavardhan B V}
\author[1, *]{Mohit Kumar Jolly}

\affil[1]{Department of Bioengineering, Indian Institute of Science, Bengaluru, Karnataka, India - 560012}
\affil[2]{IISc Mathematics Initiative, Indian Institute of Science,  Bengaluru, Karnataka, India - 560012}
\affil[*]{Corresponding Author, Email: mkjolly@iisc.ac.in}

\date{} 

\begin{document}

\maketitle

\vspace{-1.1cm} 


\begin{abstract}
The emergent dynamics of complex gene regulatory networks govern various cellular processes. However, understanding these dynamics is challenging due to the difficulty of parameterizing the computational models for these networks, especially as the network size increases. Here,  we introduce a simulation library, Gene Regulatory Interaction Network Simulator (GRiNS), to address these challenges. GRiNS integrates popular parameter-agnostic simulation frameworks, RACIPE and Boolean Ising formalism, into a single Python library capable of leveraging GPU acceleration for efficient and scalable simulations. GRiNS extends the ordinary differential equations (ODE) based RACIPE framework with a more modular design, allowing users to choose parameters, initial conditions, and time-series outputs for greater customisability and accuracy in simulations. For large networks, where ODE-based simulation formalisms do not scale well, GRiNS implements Boolean Ising formalism,  providing a simplified, coarse-grained alternative, significantly reducing the computational cost while capturing key dynamical behaviours of large regulatory networks. The documentation and installation instructions for GRiNS can be found at \href{https://moltenecdysone09.github.io/GRiNS/}{https://moltenecdysone09.github.io/GRiNS/}.
\vspace{0.1cm}
\end{abstract}

\section*{Keywords:}
Gene Regulatory Networks, Network Dynamics, Parameter-Agnostic Simulation, Ising Boolean Formalism, Random Circuit Perturbation (RACIPE), Systems Biology

\section{Introduction}
One of the core goals of systems biology is to understand and model complex interactions in a biological system to understand the mechanistic underpinnings of various cellular processes. The advent of high-throughput technologies capable of capturing the molecular profiles of cells has enabled us to infer causal relations between molecules governing various biological processes \cite{yuanInferringGeneRegulatory2025, ishikawaRENGEInfersGene2023, badia-i-mompelGeneRegulatoryNetwork2023}. These causal interactions between genes are typically represented using gene regulatory networks (GRNs). GRNs are network representations of the causal interactions between genes that govern their expression levels and functional activity \cite{barbutiSurveyGeneRegulatory2020}. GRNs are crucial for understanding the mechanistic details governing complex biological processes that give rise to specific expression patterns and cellular phenotypes \cite{gomez-schiavonArtModelingGene2024, gedeonNetworkTopologyInteraction2024, karlebachModellingAnalysisGene2008, bocciTheoreticalComputationalTools2023a}. As representations of the interactions between genes, GRNs provide a framework to decipher the regulatory logic and identify key interactions that determine a particular biological outcome. There are two main approaches to constructing GRNs \cite{jollyComputationalSystemsBiology2017}. The bottom-up approach employs detailed experiments to verify the interactions between genes to build these GRNs, with experiments to tease out causal relations between genes. However, while being precise, this approach is cumbersome because, with the increasing network size, it becomes complicated to accurately determine all the interactions between the genes. On the other hand, the top-down approach leverages high-throughput technologies and the vast amounts of data they generate to infer the interactions between genes. Various algorithms have been developed that utilize and integrate data from different single-cell molecular profiling techniques to infer gene regulatory networks \cite{yuanInferringGeneRegulatory2025, ishikawaRENGEInfersGene2023, badia-i-mompelGeneRegulatoryNetwork2023, wangRegVeloGeneregulatoryinformedDynamics2024}. However, unlike the bottom-up approach, such inferred GRNs often have limited predictive potential due to missing or inaccurate regulatory links \cite{pratapaBenchmarkingAlgorithmsGene2020}.

When it comes to understanding the dynamics of such predicted GRNs, most of these methods suffer from a common challenge of being unable to accurately predict the interaction functions and their parameters \cite{pratapaBenchmarkingAlgorithmsGene2020, markuTimeseriesTranscriptomicsGene2023}. Given the uncertainty in data due to biological noise and technical limitations, and considering the sizes of the large networks that govern these interactions, it becomes challenging to parameterize these networks accurately \cite{stadterBenchmarkingNumericalIntegration2021, frohlichFidesReliableTrustregion2022}. Having simulation frameworks that are parameter agnostic and focus on the dynamics and steady states of a given network structure becomes important in such cases. These approaches do not depend on specific parameter sets to explain the behavior of GRNs. An advantage of such approaches is that, even in the absence of parameters, having qualitative estimates of the general ranges of the parameters is sufficient for getting an idea about the dynamics of the network in the parameter range.

RAndom CIrcuit PErturbation (RACIPE) is one such framework that tries to identify the possible phenotypic space of a network given its topology \cite{huangInterrogatingTopologicalRobustness2017, huangRACIPEComputationalTool2018}. It samples parameters over predefined ranges and simulates them over multiple initial conditions.  The tool can capture a network’s steady states by randomly sampling the parameters and simulating them over multiple initial conditions. This approach helps one to understand the possible types of dynamics and steady states a network can show, even when the mathematical model’s precise kinetic parameters are absent.
Additionally, due to the random sampling of parameters and initial conditions, the RACIPE tool mimics the wide range of variability observed in biology, giving us a more realistic view of the network’s behavior compared to when we use precise parameters, which, more often than not, are very context-specific and may not represent the actual behavior of the network in a different scenario. As RACIPE automates the entire pipeline of model building and its subsequent simulation according to the inputs provided by the user, it proves to be a suitable tool for analyzing the dynamics of GRNs. It has been used to model and explain various cellular processes, including cell fate decisions, phenotypic heterogeneity, and transitions such as epithelial–mesenchymal plasticity across diverse biological contexts \cite{deyPotentialLandscapesBifurcations2021, heIdentifyingKeyFactors2023, gedeonNetworkTopologyInteraction2024}.

Another such parameter-agnostic method is the Ising boolean simulation framework \cite{liYeastCellcycleNetwork2004, barbutiEncodingBooleanNetworks2021}. This framework represents each gene as a binary variable—active or inactive— depending on the cumulative influence of the incoming active links. Since each update step is based on matrix multiplication, this method, although very crude, is suitable for simulating large networks where the system of ODEs is too large and, hence, is too slow to be practical to be simulated over thousands of parameters and initial conditions through RACIPE. It has been applied to analyze gene networks and their state transitions with minimal parameter dependence \cite{liYeastCellcycleNetwork2004, liLogBTFGeneRegulatory2023, font-closTopographyEpithelialMesenchymal2018}.

\begin{figure}[h]
    \centering
    \includegraphics[width=0.7\textwidth]{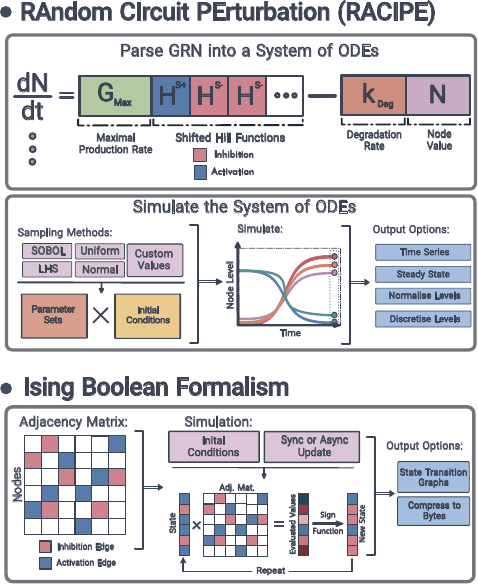}
    \caption{Overview of the simulation frameworks in GRiNS. GRiNS includes implementations of Random Circuit Perturbation (RACIPE) for continuous ODE-based modeling and Ising Boolean formalism for discrete-state simulations.}
    \label{fig:grins_fig}
\end{figure}

Because both these methods depend heavily on matrix-based operations, they are ideal candidates for GPU implementation. GPUs are suitable for matrix-based operations because they can parallelize the computation, giving dramatic speed-ups compared to the same computations run on a CPU. The Python library \textit{Jax’s} efficient array-oriented numerical computation functions provide an excellent foundation for our simulation library \cite{bradburyJAXComposableTransformations2018}. RACIPE, in particular, requires GPU-based differential equation solver implementations provided by the \textit{diffrax} \cite{kidgerNeuralDifferentialEquations2022a} library, which is built on top of \textit{Jax}. Utilizing these, we introduce a simulation library for GRNs that exploits the speed of GPUs over CPUs for matrix operations, enabling rapid and scalable simulations of GRNs (Fig. \ref{fig:grins_fig}). Our library’s modular design, written in an easily accessible language like Python, offers more options for the users to tweak the simulations according to their needs. 

\section{Methods}

\subsection{RAndom CIrcuit PErturbation (RACIPE)}
RACIPE is a GRN modeling framework designed to sample the steady-state repertoire of a GRN purely based on its topological structure \cite{huangRACIPEComputationalTool2018}. It does this by generating a system of coupled ordinary differential equations (ODEs) to represent the interactions of the genes and then simulating these equations over multiple randomly sampled parameter sets and initial conditions. By simulating a parameterized system of  ODE over the initial conditions, RACIPE can uncover the possible set of steady states. When this process is carried out on all the parameter sets, RACIPE can map out the possible phenotypic outcomes of a network even in the absence of specific parameter sets.

In the following sections, we describe the methodology of RACIPE and how it constructs the differential equation models from network topologies and systematically samples parameters within biologically relevant ranges. We will then describe the simulation methodology of the parameterized ODEs over the initial conditions and the subsequent analysis, which can be done to identify robust, steady states to uncover the dynamic landscape of gene regulatory networks.

\subsubsection{Parsing GRNs to Construct System of ODEs}
RACIPE models a signed and directed GRN as a system of coupled ODEs. For a gene, T, in the GRN, the ODE describing the change in its expression value as a function of the input nodes is given by:

\begin{equation}
    \begin{split}
        \frac{dT}{dt} = G_T * \prod_i \frac{H^{S}(P_i, {P_i}^{0}_{T}, n_{P_iT}, \lambda_{P_iT})}{\lambda_{P_i, T}} * \prod_j H^{S}(N_j, {N_i}^{0}_{T}, n_{N_jT}, \lambda_{N_jT}) - k_T*T
    \end{split}
    \label{eq:racipenode_eq}
\end{equation}

Where,
\begin{equation}
	\begin{split}
	    H^{S}(B, B^0_A, n_{BA}, \lambda_{BA}) = \frac{{B^0_A}^{n_{B,A}}}{{B^0_A}^{n_{BA}} + B^{n_{BA}}} + \lambda_{BA} * \frac{B^{n_{BA}}}{{B^0_A}^{n_{BA}} + B^{n_{BA}}}
	\end{split}
    \label{eq:shiftedhill_eq}
\end{equation}

$G_T$ in \Cref{eq:racipenode_eq} refers to the maximal expression value of the node T, $H^s$ is a modified form of Hill’s equation called the Shifted Hill's equation, as given in \Cref{eq:shiftedhill_eq}. Each Shifted Hill's  \Cref{eq:shiftedhill_eq} represents the effect of an upstream incoming node on T. It has the threshold parameter $B^0$, $n_{BA}$ is the hill’s coefficient, and $\lambda_{BA}$ is the fold change parameter representing the fold change in the expression of A brought about by the effect of the edge from node B. $P_i$, and $N_j$ refers to the values of the input activating and inhibiting nodes respectively. All the hill’s terms of the incoming edges are multiplied with the maximal expression value $G_T$ to get a scaled value of T’s expression, the production term of the ODE. Additionally, $k_T$ is the degradation rate of the node and is multiplied by the value of T to get the degradation term of the equation.

The current parser supports only signed and directed GRNs, i.e., it only recognizes activation and inhibition links when constructing the system of ODEs. Once generated, the ODE system is written into a Python file formatted for compatibility with the Diffrax simulation library. Users can modify these functions to incorporate custom dynamics, as the subsequent simulation steps do not depend on the original RACIPE ODE structure given above. 

\subsubsection{Sampling Parameters and Initial Conditions}
Since RACIPE’s primary object is to capture all possible steady states of a given GRN, the parameter sampling strategy must eliminate any biases that may skew the result. The number of parameters to be sampled for a network with N nodes and E edges is 2N + 3E. The term 2N represents the maximal production and degradation rate parameters that must be sampled. The 3E term represents the threshold, the hill’s coefficient, and the fold change parameters that need to be sampled for each of the edges present in the network. \Cref{tab:parameters_table} gives the default ranges of the parameter values over which the sampling is done. Th

\begin{table}[h]
    \centering
    \renewcommand{\arraystretch}{1.3} 
    \begin{tabular}{|l|c|c|}
        \hline
        \textbf{Parameters} & \textbf{Minimum} & \textbf{Maximum} \\
        \hline
        Production Rate (G) & 1 & 100 \\
        \hline
        Degradation Rate (k) & 0.1 & 1 \\
        \hline
        Fold Change (Inhibition $\lambda$) & 0.01 & 1 \\
        \hline
        Fold Change (Activation $\lambda$) & 1 & 100 \\
        \hline
        Hill Coefficient (n) & 1 & 6 \\
        \hline
        Threshold & \multicolumn{2}{|c|}{The ranges depend on the in-degree - half functional rule} \\
        \hline
    \end{tabular}
    \caption{Default parameter ranges used by RACIPE.}
    \label{tab:parameters_table}
\end{table}

The inhibition fold change parameter is sampled in the inverse range, i.e., between the inverse of the maximum value and the inverse of the minimum value. Following this, the inverse of all the sampled values is taken. This process shifts the mean of the distribution from 0.5 to 0.02 (in the case of default ranges), which ensures that both weak and strong inhibitory fold change parameters are sampled in a balanced manner.

Another aspect to consider to eliminate biased parameters of nodes is the threshold values assigned to the edges. For nodes with incoming edges, the threshold values need to be sampled to be within the minimal and maximal steady-state expression ranges of the nodes from which they originate. Otherwise, a deviation from this would mean that an edge could be always active or always inactive, biasing the simulations. The half-functional rule is employed to correct this bias, and it ensures that the threshold values are chosen between 0.02 and 1.98 times the median steady-state expression level of the upstream node. Suppose other genes regulate the upstream node itself. In that case, its steady-state expression distribution is determined by simulating its steady-state levels based on the regulatory parameters of its incoming nodes. The median expression of the upstream node is then used to define the appropriate range for threshold sampling. This ensures that the threshold values are chosen so that each regulator has a probability of being active at roughly 50\% across all the simulations.

The values are sampled from the respective node’s minimum and maximum expression values for the initial conditions. Our framework supports multiple sampling methods—including Sobol (default), Latin hypercube, uniform, log-uniform, normal, and log-normal distributions—and even allows mixing different distributions for different parameters or initial conditions \cite{virtanenSciPy10Fundamental2020}. This flexibility lets users target specific regions of parameters or the initial condition space. A parameter range file is also generated for easy customization, on which the parameters can be regenerated for subsequent simulations. Similarly, initial conditions can also be regenerated based on the updated ranges.

\subsubsection{Simulating the ODE system and processing the results}
Once the parameters and initial conditions are generated after the GRN is parsed to create the ODE file, the library provides functions for the simulation of the ODE system for all combinations of parameter sets and initial condition values. A function is also provided to simulate the ODE system for a single combination of parameters and initial conditions. The simulations are usually done using the \textit{vmap} function of the \textit{Jax} library, and the user can control the batch size for cases where the VRAM is insufficient to run large-scale simulations. The user can control the tolerances (relative and absolute), the start time point (by default is zero as we would be dealing with initial conditions), and the end time point of the simulations, after which the simulation will terminate even though the steady-state may not have been reached. By default, the simulations terminate once the steady state condition is reached, which is determined by the tolerance values. The resulting data—comprising node values, termination times, a steady-state indicator, and identifiers for each parameter–initial condition pair—is stored in a data frame as a Parquet file on the disk.

Setting the end time point sufficiently high is important to allow most combinations to reach a steady state, which the steady state flag column of the solution data frame can track. The implementation also supports recording gene expression levels at custom time points for time-series analyses. However, we recommend first determining the steady-state range and then deciding on the optimal spacing of the time points and their range to capture the relevant dynamics.

Since parameters can vary across each simulation instance, a normalization method is necessary to compare steady-state values. Our library provides functions that normalize node expression by the maximal expression-to-degradation ratio (G/k) – the highest possible expression level – resulting in outputs scaled between 0 and 1, with 1 being the maximal expression of the node. Additionally, nodes can be discretized into levels based on the global distribution of normalized G/k values, offering a standardized approach to processing simulation outputs. These functions are compatible with both the steady-state and time-series simulation results.

\subsection{Boolean Ising Formalism}
Although RACIPE formalism is an excellent framework for understanding the possible dynamics emergent from a GRN, computational cost becomes prohibitive for large networks due to the increasing number of parameters and longer simulation times. Boolean Ising formalism, also referred to as threshold boolean formalism, provides a simpler, albeit coarse-grained approach for simulating large networks \cite{font-closTopographyEpithelialMesenchymal2018, barbutiEncodingBooleanNetworks2021, barbutiSurveyGeneRegulatory2020}. In this approach, the state of the variable is represented by a discrete variable indicating the on or off state of the gene. The system starts with an initial condition ($s_0$), and the subsequent state ($s_{t+1}$) is obtained by the previous state’s ($s_t$) matrix product with the network's adjacency matrix (A), where -1 represents inhibitory interactions, 1 represents activating interactions, and 0 indicates no connection. Once the state is multiplied with the adjacency matrix, the resultant vector is then converted to 1s and -1s (or 0s depending on the flip values provided) according to the rule given below:

\begin{equation}
    s_i(t+1) =
    \begin{cases} 
        1, & \sum\limits_{j} J_{ij} s_j(t) > 0 \\
        s_i(t), & \sum\limits_{j} J_{ij} s_j(t) = 0 \\
        -1, & \sum\limits_{j} J_{ij} s_j(t) < 0
    \end{cases}
    \label{eq:ising_eq}
\end{equation}

The system can be updated using either a synchronous update, where all the nodes are updated simultaneously, or an asynchronous update, where one randomly chosen node is updated at each step. The user can set the number of simulation steps, define initial conditions (or use randomly generated ones by default), and specify a custom update order for asynchronous mode. The simulation results are stored in a dataframe with columns for the binary node states (0 or 1), the simulation step number, and the corresponding initial condition number. Optionally, the node values at each step can be stored as bytes to conserve disk space and later unpacked during analysis.

\section{Case Studies}

\subsection{Comparison of the steady state dynamics of Toggle Switch and a Toggle Switch with Self-activation}
\begin{figure}[htbp]
    \centering
    \includegraphics[width=0.9\textwidth]{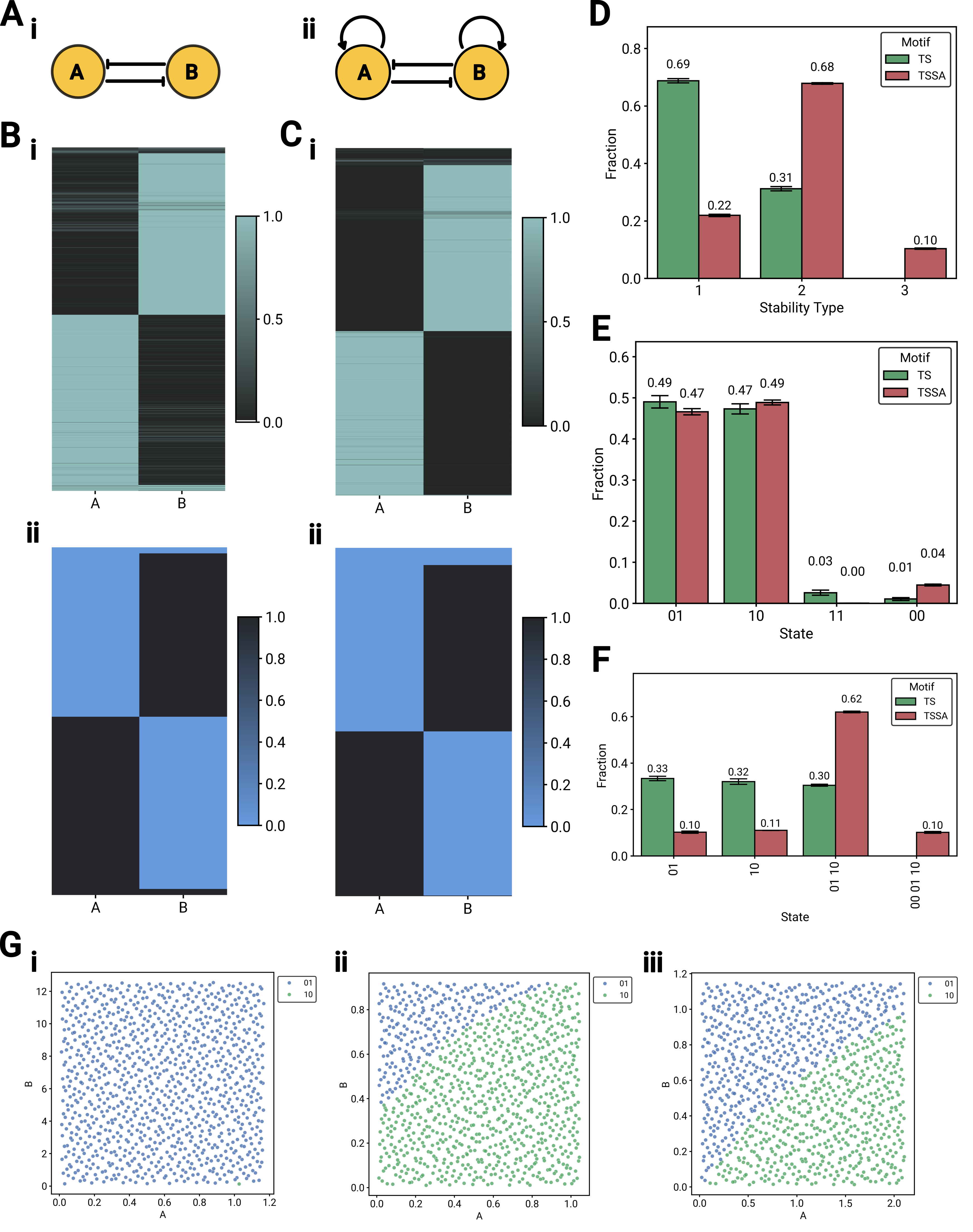}
    \caption{\textbf{Comparison of toggle switch (TS) and toggle switch with self-activation (TSSA) motifs using the GRiNS framework.} \textbf{A)} The toggle switch (TS) \textbf{(i)} and the toggle switch with self-activation \textbf{(ii)}. \textbf{B)} Sample of expression profiles of TS nodes. \textbf{i)} G/k normalised expression \textbf{ii)} Discretised levels of the normalised expression values. \textbf{C)} Sample of expression profiles of TSSA nodes. \textbf{i)} G/k normalised expression \textbf{ii)} Discretised levels of the normalised expression values. \textbf{D)} Distribution of the parameters according to their multistability for TS and TSSA. \textbf{E)} Distributions of the steady states of TS and TSSA. \textbf{F)} Comparison of the multistable steady state frequencies between TS and TSSA. G) Initial condition (G/k normalised) mapping for three different bistable parameter sets.}
    \label{fig:cs1_fig}
\end{figure}

To demonstrate the utility of our simulation framework, we present a comparative analysis of the toggle switch (TS) (\Cref{fig:cs1_fig}A, i) and its variant with self-activation on both nodes (TSSA) (\Cref{fig:cs1_fig}A, ii). The TS motif is commonly found in gene regulatory networks and is known for exhibiting bistability, enabling cells to maintain expression states that let them commit to distinct fates \cite{heinaniemiGenepairExpressionSignatures2013, huangBifurcationDynamicsLineagecommitment2007, mojtahediCellFateDecision2016}. The mutual inhibition between the two genes results in antagonistic expression profiles—an essential property for cell fate commitment—as it prevents promiscuous co-expression of the genes belonging to both the cell types \cite{guantesMultistableDecisionSwitches2008}. 

We use the RACIPE implementation in the GRiNS library to simulate these motifs and contrast the effect of adding self-activating loops to the TS motif in terms of steady states, the distribution of multistable states, and the normalization and analysis of steady-state data across different parameter sets and initial conditions. We simulated the motifs over 10,000 parameter sets and 1,000 initial conditions, run in triplicate. Sobol sampling was used to generate both the parameters and the initial conditions. During the parameter and initial condition generation step, the function also parses the GRN topology file to generate a Python file containing the ODE representation of the GRN. The ODE system follows the structure of the RACIPE formalism laid out in the Methods section. Once generated, the ODe system file is saved in a user-specified folder named after the input topology file. The user can then also manipulate the system ODE file, providing flexibility to introduce custom functions in addition to the default shifted Hill functions used by RACIPE. In such cases, however, the generated parameters may not be compatible with simulations, and a custom parameter file would be required. The rationale for running simulations in triplicate is to ensure sufficient sampling of parameters and initial conditions from a given topology file. Since the RACIPE framework is parameter-agnostic, with its primary purpose being the identification of steady-states, their types, and distributions, it is important to assess the granularity of parameter space sampling. Too coarse a sampling could miss rare but important states; too dense a sampling increases computation time per replicate. We deemed the number of parameters and initial conditions sufficient for the current use case.

After parameter and initial condition generation, the Python ODE file is loaded and simulated in parallel using the diffrax library's Tsit5 ODE solver. This GPU-compatible solver significantly speeds up simulations, particularly suitable for the RACIPE framework, where the same ODE system is parameterized and simulated over many independent initial conditions. As each simulation is independent of the others, this presents an opportunity to parallelize the simulations using the GPUs to speed up the simulations.

Simulation results are reported as a parquet file containing the parameter and initial condition indices, a steady-state flag (one if no steady state is reached within the user-defined time step, zero otherwise), and the steady-state values of the nodes. Because parameter values are randomly sampled, directly comparing raw steady-state values across topologies or parameter sets is not meaningful. To address this, we apply G/k normalization to scale steady-state values between 0 and 1, where 1 indicates the maximal possible expression level for a gene under the given parameter set. For both TS and TSSA motifs, the heatmaps show that most steady states involve one node expressing close to 1 (high expression) and the other close to 0 (low expression), consistent with mutually exclusive expression (\Cref{fig:cs1_fig}B, i; \Cref{fig:cs1_fig}C, i).

However, since these normalized values are still continuous, it becomes difficult to analyze and compare the state spaces of large or multiple GRNs. To address this, a function in our library allows coarse-grain expression levels into discrete "levels," simplifying comparisons across parameter sets and networks. This approach simplifies the further use of complex dimensionality reduction or clustering algorithms, making it scalable to larger GRNs. After the coarse-graining step, the steady states of TS and TSSA reduce to binary values (0s and 1s) (\Cref{fig:cs1_fig}B, ii; \Cref{fig:cs1_fig}C, ii). The binary nature of the levels reveals that both TS and TSSA motifs primarily exhibit two expression levels per node. 

The highly bimodal nature of the motif nodes is evident in their G/k normalized expression values, too, which are either very high, close to 1, or very low, close to 0. Based on the levels of the nodes for a particular parameter and initial condition set, we then assign it a state composed of the concatenated levels of the nodes. We then examined the distribution of these coarse-grained steady states for multistability. The multistability of a parameter refers to the number of unique steady states it produces when simulated across all initial conditions. Most of the sampled TS parameters were monostable, while TSSA shows an increased number of bistable and a small number of tristable parameters (\Cref{fig:cs1_fig}D). Next, when we looked at the distribution of the states themselves, consistent with expectations from the heatmaps, both motifs most frequently produce the 10 and 01 states. Although all four binary combinations (00, 01, 10, 11) are observed, TSSA exhibits a higher frequency of the 00 state than TS, which shows both 00 and 11, albeit rarely (\Cref{fig:cs1_fig}E).

Continuing with the multistability analysis of the parameter sets, we observed that TSSA frequently exhibits bistability between 01 and 10 (\Cref{fig:cs1_fig}F). At the same time, TS mainly shows monostable 01 or 10 states, with bistable 01–10 states occurring less frequently. TSSA produces tristable states that consist of the 00, 01, and 10 states, which explains the higher occurrence of tristability in TSSA compared to TS. Going beyond the mere presence of multistability, one can also examine the relative frequencies of the coarse-grained steady states for a particular parameter set across multiple initial conditions. To illustrate this, we randomly selected three bistable parameter sets that differ in the frequencies of the 01 and 10 states they produce. The plots of the initial conditions of the node values, colored by the corresponding steady-state they reach (\Cref{fig:cs1_fig}G, i-iii). This functionality—to track each initial condition and the steady state it leads to—can be combined with parameter sampling to investigate how the steady-state landscape changes as a specific parameter is varied. Since the initial conditions remain the same across parameter sweeps, the individual data points can be stacked, enabling more detailed analysis.
In conclusion, the comparison between the TS and TSSA motifs revealed that the addition of self-activation on the nodes increased the ability to show bistability between the two mutually exclusive states of 01 and 10.

\subsection{EMT Network - Comparison between RACIPE and Ising Boolean Formalisms}
\begin{figure}[htb]
    \centering
    \includegraphics[width=0.85 \textwidth]{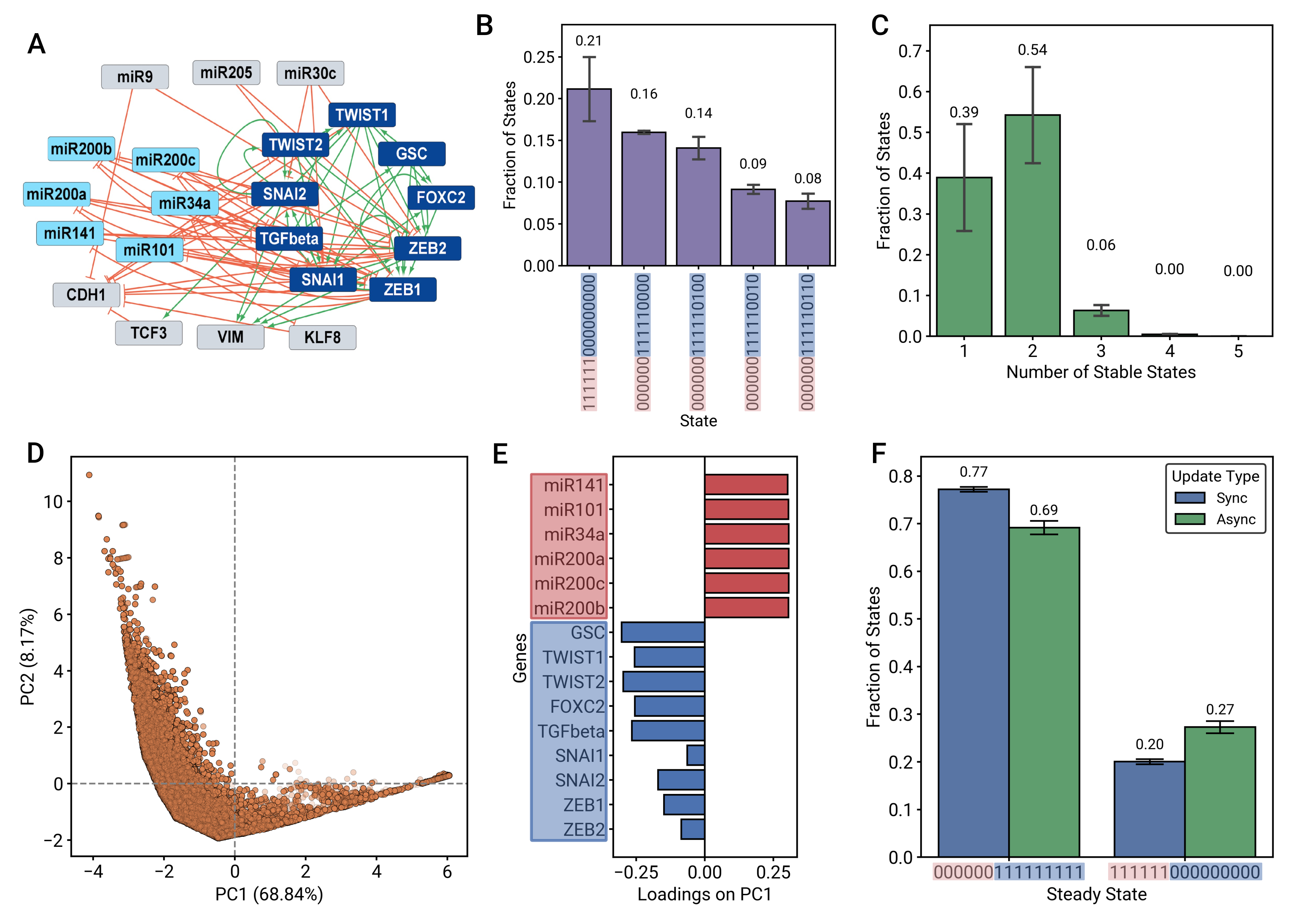}
    \caption{\textbf{Analysis of the EMT gene regulatory network using RACIPE and Ising Boolean formalisms.} \textbf{A)} Network diagram of the 22-node EMT GRN with 82 regulatory edges. Input and output nodes (marked grey) and excluded from analysis. \textbf{B)} Discretised steady-state distributions from RACIPE simulations. \textbf{C)} Distribution of multistability types from RACIPE simulations. \textbf{D)} PCA of G/k normalized steady states. \textbf{E)} PC1 loadings from the PCA. \textbf{F)} Steady state frequency distribution from synchronous and asynchronous Ising Boolean simulations.}
    \label{fig:cs2_fig}
\end{figure}

We now move on to a larger network—the EMT (epithelial-to-mesenchymal transition) network, which consists of 22 genes and 82 interactions (\Cref{fig:cs2_fig}A) \cite{font-closTopographyEpithelialMesenchymal2018}. This case study demonstrates how the same methods used for TS and TSSA can be scaled to larger networks and what additional insights can be gained from such simulations. The EMT network governs the transition of epithelial cells into mesenchymal cells, a process often dysregulated in cancer \cite{simeoneMultiverseNatureEpithelial2019, dongreNewInsightsMechanisms2019, francouEpithelialtoMesenchymalTransitionDevelopment2020}. This dysregulation can result in hybrid phenotypes that are more challenging to treat and are associated with increased resistance to therapies \cite{jollyHybridEpithelialMesenchymal2018}.
We simulated this network in triplicate, sampling 10,000 parameter sets and 100 initial conditions for each replicate. One way to assess whether the number of parameters and initial conditions chosen is sufficient is to examine the distribution of coarse-grained steady states across replicates. If the standard deviations across replicates are low, it suggests that the sampling is consistent and that results are reliable, i.e., different replicates yield similar steady-state distributions. In our case, the low variability observed confirms that the selected number of parameter sets and initial conditions adequately capture the diversity of the GRN's steady states (\Cref{fig:cs2_fig}B).

The EMT network contains several input nodes (with only outgoing edges) and output nodes (with only incoming edges). Due to how RACIPE is set up, randomly sampling initial conditions means that input nodes may be turned on or off arbitrarily without any regulation by other genes. Similarly, output nodes reflect the status of their upstream regulators. To make the analysis more focused and meaningful, we excluded the expression profiles of these input and output nodes from further analysis. An alternative way to address this issue is to modify the initial conditions of the input nodes based on the context of the simulation. For instance, one could set the input nodes connected to mesenchymal genes to high values, thereby simulating the influence of external factors on the network. In such scenarios, the limitations associated with randomly chosen initial conditions would no longer apply, and excluding these nodes from the analysis would not be necessary.

One notable observation from the coarse-grained steady-state distributions is that epithelial genes tend to be strongly co-expressed, whereas mesenchymal genes do not show consistent co-expression (\Cref{fig:cs2_fig}B). Self-inhibitory interactions in the network, like the ones present on SNAI1 and ZEB1, may partly explain this variability among mesenchymal genes. Despite the EMT network's size, most parameter sets produced majorly bistable followed by monostable outcomes, suggesting strong coordination among genes leading to a constrained state space (\Cref{fig:cs2_fig}C), a pattern consistent with previous studies  \cite{hariLowDimensionalityPhenotypic2025, huangInterrogatingTopologicalRobustness2017}. To explore these states further, we performed Principal Component Analysis (PCA) on the steady-state gene expression profiles normalized by their G/k ratios. As the coarse-grained analysis indicated, PCA results confirm that epithelial genes cluster tightly as seen by the distribution of the points on the positive side of the PC1 axis, reflecting strong co-expression (Fig. \ref{fig:cs2_fig}D). This trend is also confirmed by the PCA loadings of the first principal component: epithelial genes have similar values, while mesenchymal genes exhibit more spread in their loadings (Fig. \ref{fig:cs2_fig}E). Overall, this analysis suggests that even though coarse-graining reduces the granularity of continuous data, it still preserves essential features and makes the interpretation of large network simulations more intuitive and tractable. 

An alternative approach for handling such large networks is to use the Ising Boolean formalism. Due to its non-parametric nature and implementation via matrix multiplications, this method is faster and can quickly approximate the steady-state landscape of a network. This becomes especially important for networks like EMT or even larger ones, where the computational cost and VRAM limitations make solving large systems of ODEs impractical. We simulated the network using both synchronous and asynchronous Ising update modes. Both methods produced the same dominant states: one where epithelial genes are active and mesenchymal genes are inactive, and the other with the reverse pattern (\Cref{fig:cs2_fig}F). However, unlike RACIPE simulations, the Ising formalism failed to capture intermediate or hybrid states. This limitation stems from the structure of the formalism, since no parameters are assigned to nodes or edges, the nuances that give rise to hybrid states are not preserved. Regarding the differences between the two update rules, synchronous updates produce deterministic outcomes and do not account for multistability. On the other hand, asynchronous updates can lead to different outcomes from the same initial conditions due to the randomness in update sequences. Consequently, the variation in state frequencies between the two modes results in the asynchronous mode showing a greater number of unique states as compared to the synchronous mode, this also results in a lower value of the fraction of the most frequent state seen as the state distribution is relatively flatter, especially for the larger networks.

When comparing the results from RACIPE and the Ising formalism, both approaches consistently showed a higher frequency of the epithelial-low/mesenchymal-high state (although the states where mesenchymal genes are on do not tend to have all of them being on together) than the opposite of epithelial-high/mesenchymal-low states (\Cref{fig:cs2_fig}B; \Cref{fig:cs2_fig}F). Overall, the ODE-based approach offers richer data and enables more detailed analyses than the Ising formalism. However, simpler models such as the Ising Boolean formalism for large-scale exploratory studies can still offer valuable approximations of a network's state space composition, especially when many networks need to be analyzed and compared. The Ising formalism can be a good approximation of a network's dynamics and act as an initial step toward understanding the network during analysis.


\section{Conclusion}
GRiNS uses parameter-agnostic simulation frameworks and sampling strategies to model GRNs and capture their dynamics under various conditions. Implementing these methods in an accessible language like Python coupled with modular functions allows users to customize their simulations and explore the dynamic properties of GRNs in a standardized and reproducible manner. Thanks to its GPU-based simulation, GRiNS can be easily integrated into machine learning pipelines for cases like network inference, where unbiased simulation methods are important to determine and evaluate the predicted networks. The documentation and installation instructions for GRiNS can be found at \href{https://moltenecdysone09.github.io/GRiNS/}{https://moltenecdysone09.github.io/GRiNS/}.

\section{Availability and Requirements}

\begin{itemize}
    \item \textbf{Project name:} Gene Regulatory Interaction Network Simulator (GRiNS)
    \item \textbf{Project home page:} \href{https://moltenecdysone09.github.io/GRiNS/}{https://moltenecdysone09.github.io/GRiNS/}
    \item \textbf{Operating system(s):} Platform Independent
    \item \textbf{Programming language:} \textit{Python3}
    \item \textbf{License:} GNU GPL-3.0 License
\end{itemize}

\section*{Competing Interests}
The authors declare no competing interests.

\section*{Author Contributions Statement}
M.K.J. and P.H. conceived the idea of the simulation package.  P.H. and H.B.V. wrote and developed the package. M.K.J. and P.H. wrote and reviewed the manuscript

\section*{Acknowledgments}
M.K.J. received support from Param Hansa Philanthropies. P.H. and H.B.V. were supported by their respective Prime Minister’s Research Fellowships (PMRF) awarded by the Government of India.

\bibliographystyle{unsrt}
\bibliography{GRINS}

\end{document}